\documentclass[a4paper]{jpconf}

\usepackage{epsfig}
\usepackage{latexsym}
\usepackage{graphics}
\usepackage{graphicx}
\usepackage{amssymb}
\usepackage{epstopdf}
\usepackage{subfigure}
\usepackage{collref}

\begin{document}

\title{Aspects of production and kinetic decoupling of non-thermal dark matter}

\author{G. Arcadi $^1$ and P. Ullio $^1$}

\address{$^1$ SISSA,
              Via Bonomea 265, I-34136 Trieste, Italy and\\
              Istituto Nazionale di Fisica Nucleare,
              Sezione di Trieste, I-34136 Trieste, Italy}
              
\ead{arcadi@sissa.it, ullio@sissa.it}

\begin{abstract}
 We reconsider non-thermal production of WIMP dark matter in a systematic way and using
a numerical code for accurate computations of dark matter relic densities. Candidates
with large pair annihilation rates are favored, suggesting a connection with the
anomalies in the lepton cosmic-ray ßux detected by Pamela and Fermi. Focussing on
supersymmetric models we will consider the impact of non-thermal production on the
preferred mass scale for dark matter neutralinos. We have also developed a new formalism
to solve the BoltzmannÕs equation for a system of coannihilating species without assuming
kinetic equilibrium and applied it to the case of pure Winos.
\end{abstract}

\section{General framework}

In this work we present a numerical code, now added to the package DARKSUSY \cite{Gondolo:2004sc}, aimed to determine the dark matter relic density when this is produced by the decay of of heavy long lived states with weak enough interactions with ordinary matter to be out-of-equilibrium also in the early stages of the history of the Universe. These states, referred for simplicity as cosmological moduli $X_i$, dominate the early stages of the history of Universe and lately decay into DM as well as SM particles with a sharp increase of entropy density. When this decay occurs at lower temperatures than the standard freeze-out temperature ${T}_{t.f.o.}$ this non-thermal component dominates the dark matter relic density being the component already present in the thermal bath diluted by the entropy injection. Starting from the simplest case of only one decaying field without referring a definite particle physics scenario we have numerically recovered an interesting approximate trend occuring when, during the production process the dark matter pair annihilation rate $\Gamma = n_\chi  \langle \sigma v \rangle$ is larger than the expansion rate $H$. In such a case pair annihilations are very efficient and instantaneously decrease the number density of $\chi$ to the critical density level corresponding to $\Gamma \simeq H$ when the annihilations stop. The non-thermal relic density is given by a simple scaling law as a function of the one obtained by assuming the DM candidate a thermal relic freezing-out at $T_{t.f.o.}$.
\begin{equation}
  \label{eq:ntc}
  \Omega_\chi^{NT} h^2 \simeq \frac{{T}_{t.f.o.}}{T_{\rm RH}}\, \Omega_\chi^{T} h^2
\end{equation}
with $T_{\rm RH}$ is a characteristic temperature, numerically determined by the code, identifying the end of the decay process of the cosmological moduli and that can be analytically estimated by the condition $\Gamma=H(T_{RH})$ where $\Gamma$ is the decay rate of the heavy field. 
Notice that a DM candidate $\chi$ whose thermal relic density is small compared to the DM density because the annihilation rate is too large, may become a viable dark matter candidate for
an appropriate value of $T_{\rm RH}$.  This suggest an intresting connection with anomalies detected by Pamela and Fermi which could be explained by an annihilating cross section 2-3 orders of magnitude larger than the thermal one. 

\section{Application to the MSSM}

Let's now consider a definite particle physics scenario, i.e. the MSSM with neutralino Dark Matter.
\begin{figure}[t]
\begin{center}
\begin{minipage}[htbp]{7cm}
 \includegraphics[width=7 cm, height= 5 cm, angle=360]{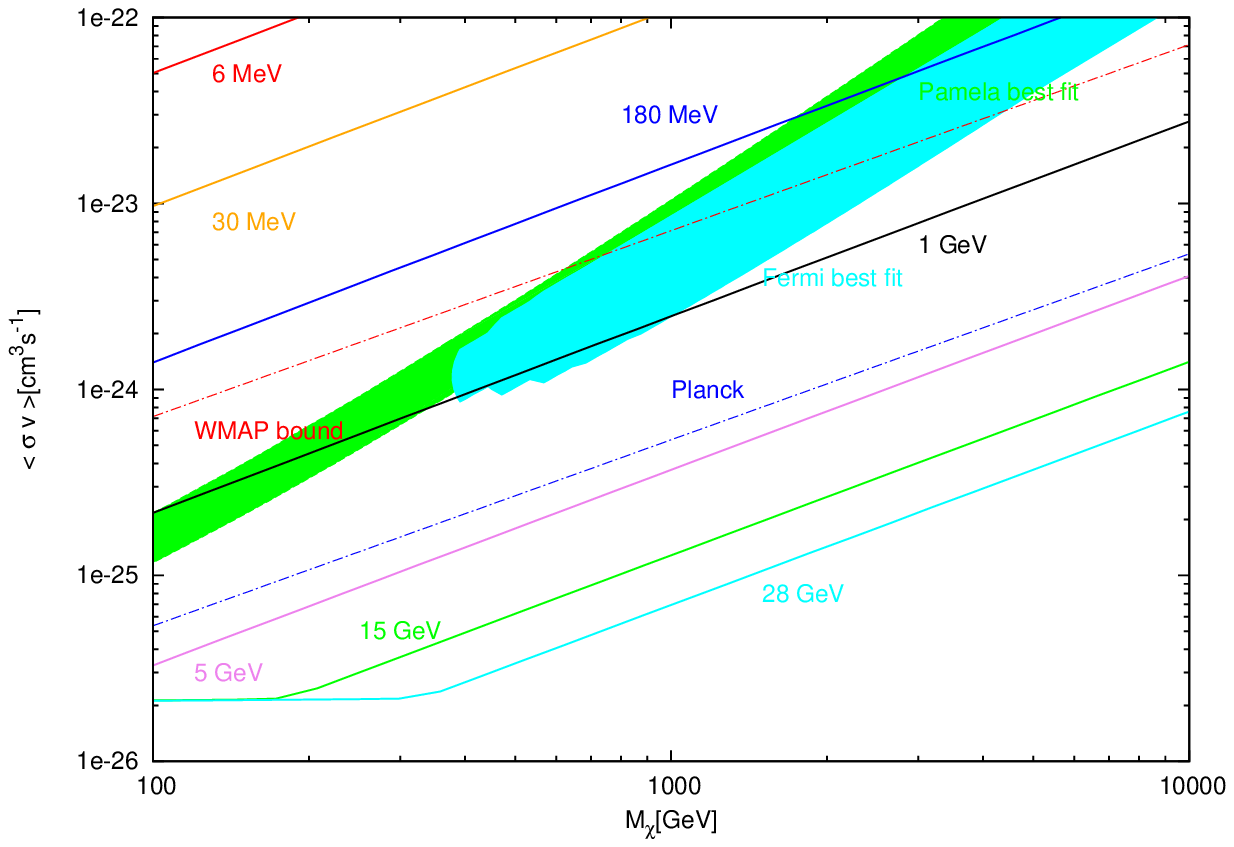}
\end{minipage}\\
\end{center}
 \begin{minipage}[htb]{7cm}
   \includegraphics[width=7 cm, height= 5 cm, angle=360]{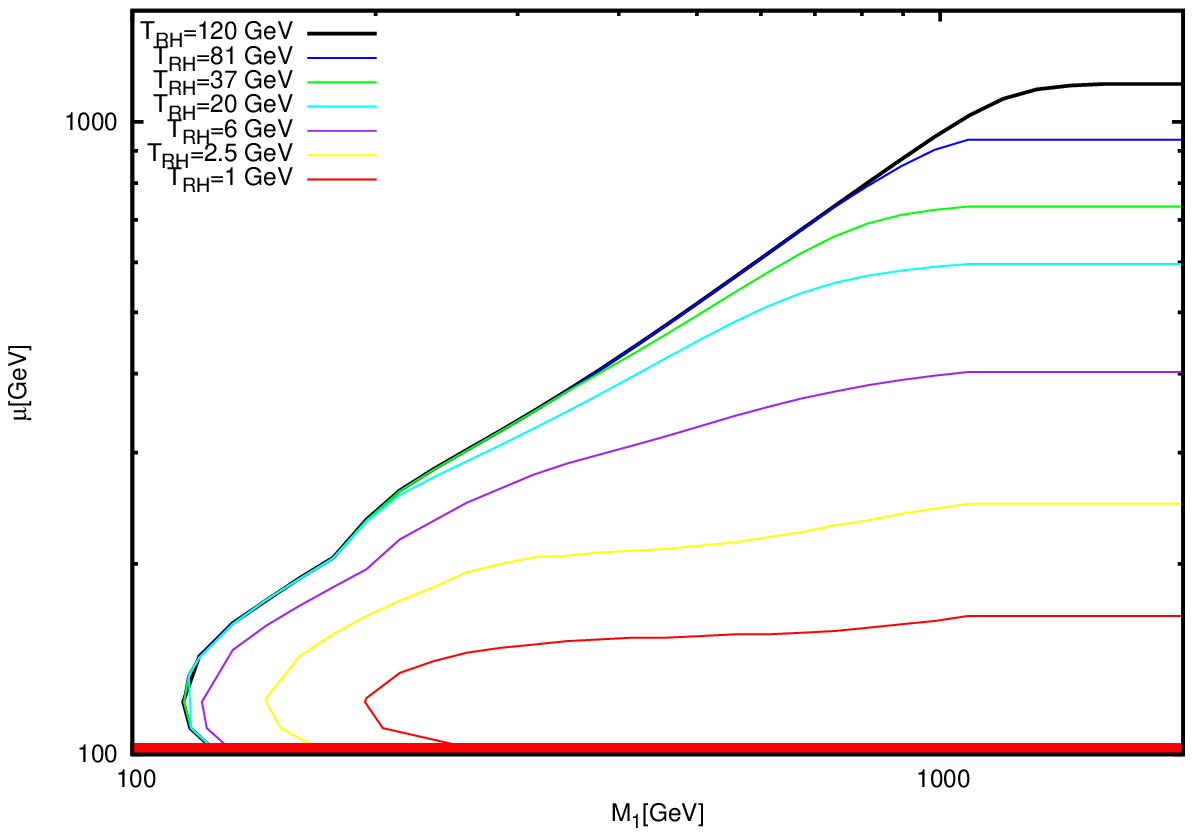}
 \end{minipage}
\hspace{2 cm}
 \begin{minipage}[htb]{7cm}
   \includegraphics[width=7 cm, height= 5 cm, angle=360]{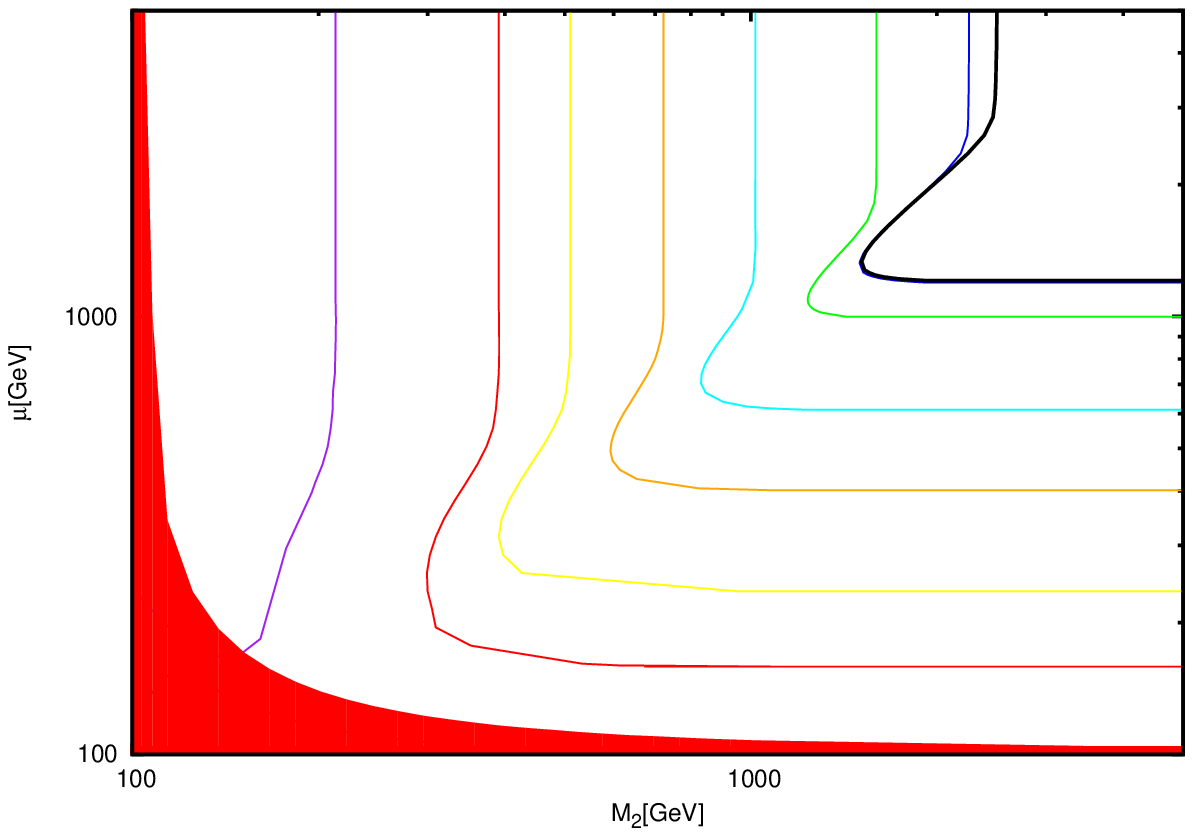}
 \end{minipage}
 \caption{Upper panel: Models fitting the cosmological value of the DM relic density in the plane $(m_{\chi}, \langle \sigma v \rangle)$. Colored regions are Pamela and Fermi best fit. The lines WMAP and Planck represent the current bound on the annihilation cross section from CMB and the Planck projected sensitivity. Lower panels:  Curves of the cosmological relic density in
  two-dimensional slices of the $M_1$, $M_2$, $\mu$ parameter space,for a few values of $T_{\rm RH}$.  The filled areas correspond to regions violating the LEP lower bound on the chargino mass.}
 \label{fig:fig2}
\end{figure}   
In the aim of determining the mass DM mass scale favored by non-thermal production we restrict to a class of models referred as split supersymmetry \cite{ArkaniHamed:2004yi,Giudice:2004tc}, characterized by very high mass scale for the scalar superpartners. In this way we could compute the non-thermal relic density for several values of the reheating temperature as a function of the Bino and the Wino mass parameters $M_1$ and $M_2$, and of the Higgs superfield parameter $\mu$.

The scenario depicted in the previous section is fitted by Wino-like or Higgsino-like dark matter. As reported of in fig.~(\ref{fig:fig2}) the cosmological value of the DM relic density can be achieved for masses of the order of one hundred GeV, contrary to the thermal case where Wino and Higgsino masses in the TeV range are required.

\section{Kinetic equilibrium and decoupling}

The results presented until now are obtained under the assumption that DM is in kinetic equilibrium with the SM states of the thermal bath along all its production process. In case that DM is part of a system of coannihilating particles, this allows to describe it by a unique Boltzmann equation for sum of their number densities with an effective annihilation cross section defined as a weighted sum of the thermally averaged cross sections of the coannihilating particles.  
The validity of the kinetic equilibrium assumption is typically guaranteed by invoking a crossing symmetry among annihilations and scattering with SM particles. This assumption may not hold in all the models. Furthermore the picture is more complicated in presence of non-thermal production. First of all DM produced out-of-equilibrium at relativistic energies and thermalization with the heat but may not occur; secondly non-thermal freeze-out may occur at low temperatures,  where the scatterings are less efficient. 

\begin{figure}[htbp]
\begin{minipage}[htbp]{7cm}
\includegraphics[width=7 cm, height= 5 cm, angle=360]{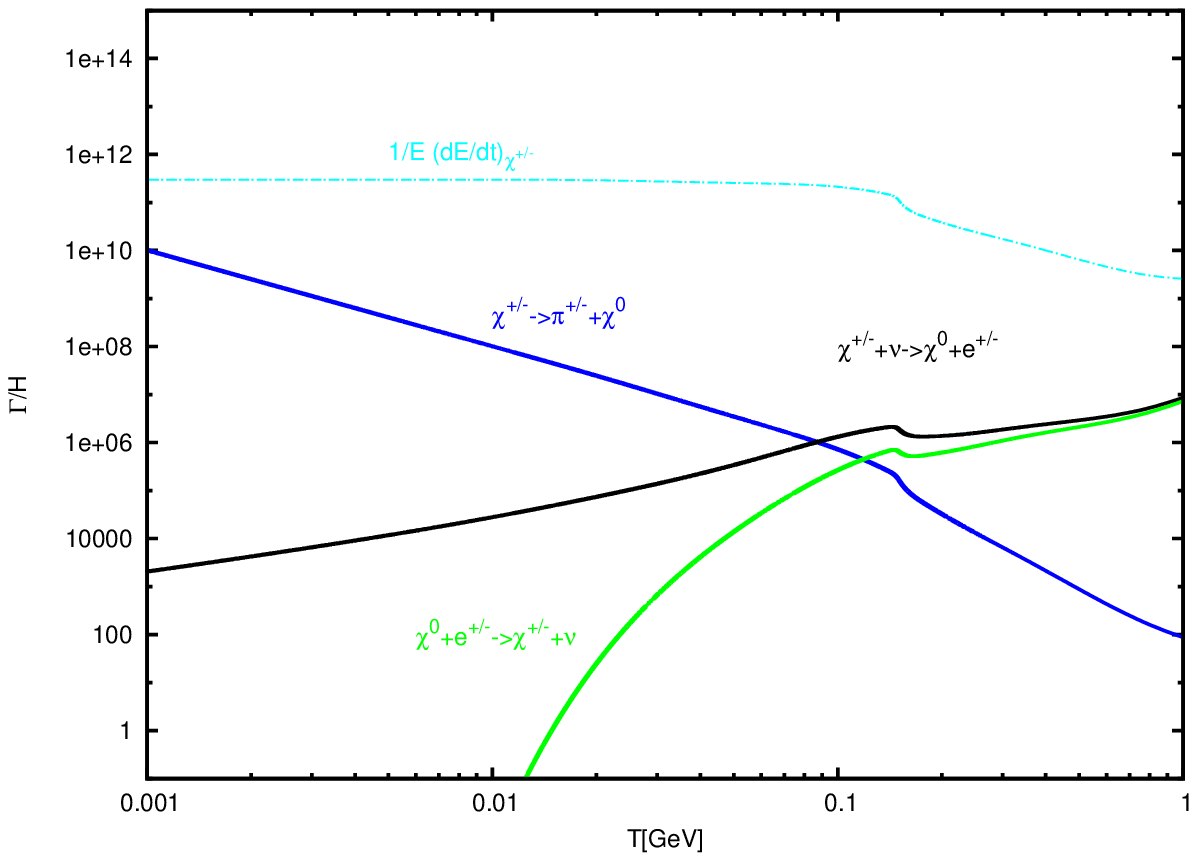}
\end{minipage}
\begin{minipage}[htbp]{7cm}
\includegraphics[width=7 cm, height= 5 cm, angle=360]{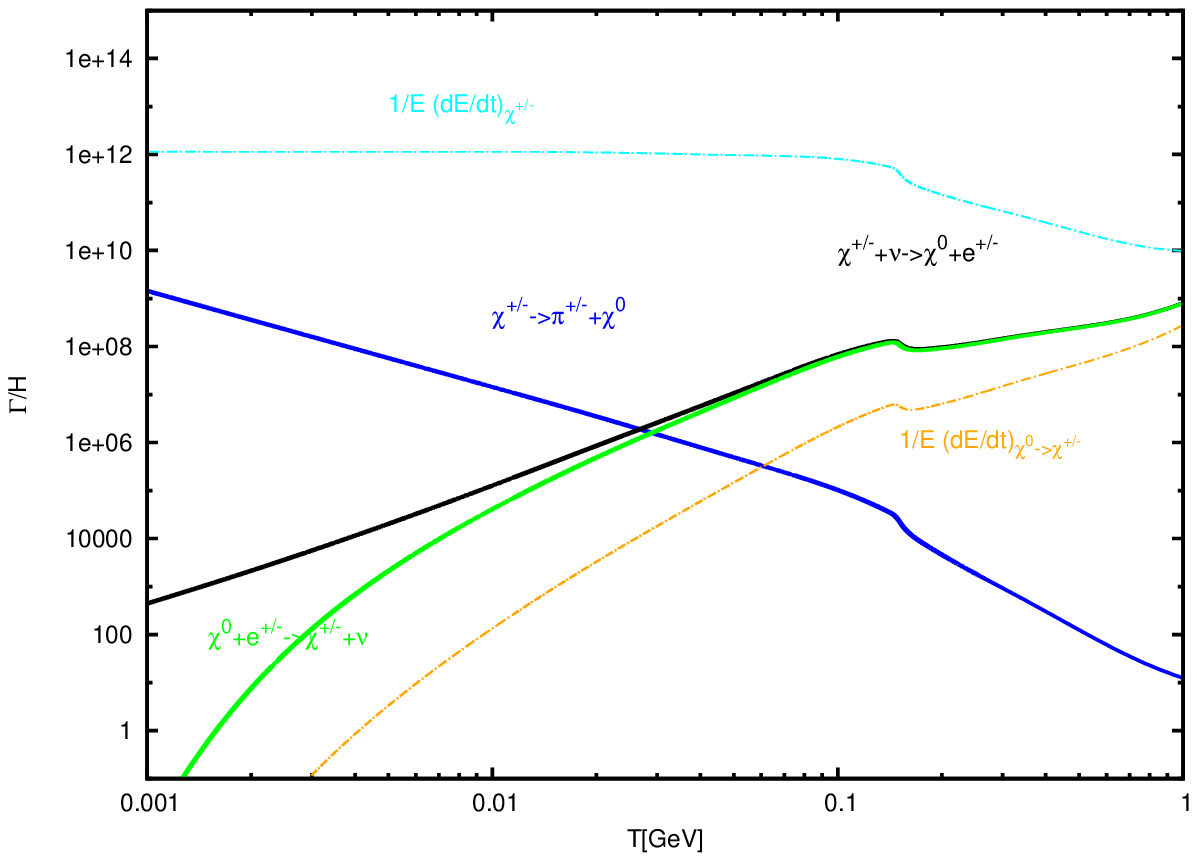}
\end{minipage}  
\caption{Ratios between the effective rate of energy loss rate $-1/E \cdot dE/dt$ (dashed lines), or of the scattering/decay rate $\Gamma$ (solid lines), to the 
Universe expansion rate $H$, for a few processes involving charginos and neutralinos. The panels refer to a G2-MSSM DM model with 
$m_\chi=103.5$~GeV and  ${T}_{RH}=100$~MeV; the plot on the
left hand-side refers to non-relativistic particles, $E/m_\chi = 1.005$, while that on the right-hand side corresponds to a sample relativistic case, $E/m_\chi =10$.    
}
\label{fig:energyloss}
\end{figure}

\begin{figure}[htbp]
 \begin{minipage}[htb]{7cm}
   \includegraphics[width=7 cm, height= 5 cm, angle=360]{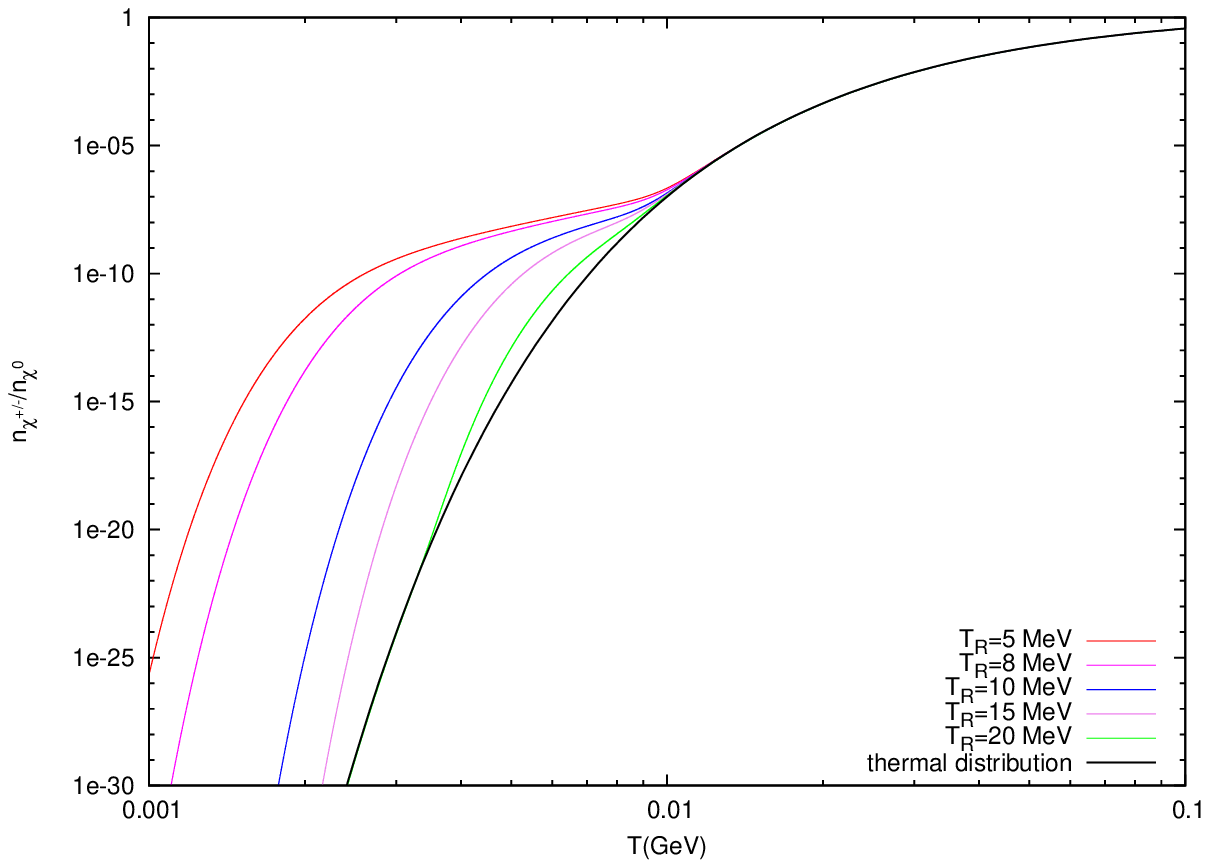}
 \end{minipage}
 \ \hspace{3mm} \
 \begin{minipage}[htb]{7cm}
   \includegraphics[width=7cm, height= 5 cm, angle=360]{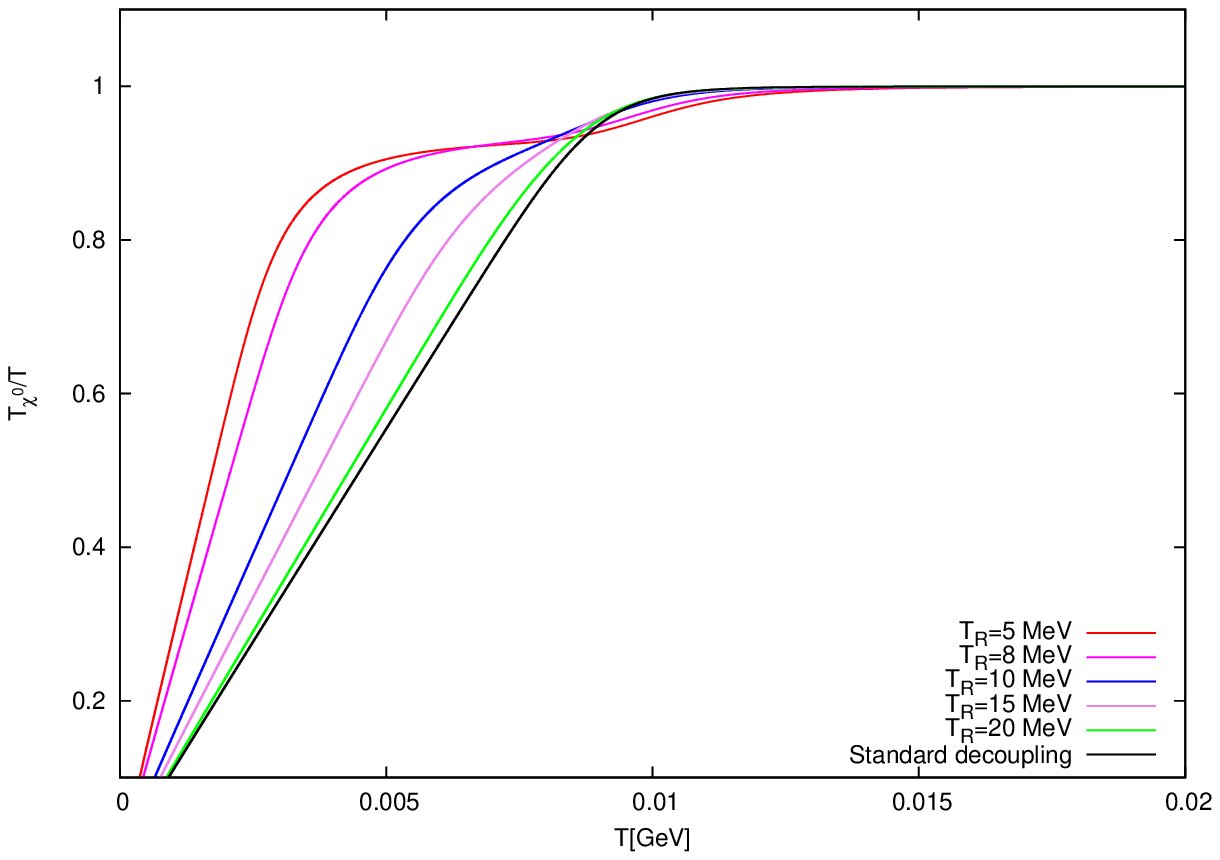}
 \end{minipage}
\caption{Left panel: ratio of the chargino number density over the neutralino number density for several values of ${T}_{RH}$. 
Right panel: Ratio $T_{{\chi}_{0}}/T$ as function of
 the temperature of the thermal bath for same values of $T_{RH}$. Plots are obtained for a Wino with mass equal to 200~GeV, however results depend
 only on the chargino-neutralino mass splitting which is about 160~MeV in the scenario under consideration.}
\label{fig:Tsever}
\end{figure}

How to address this issue varies from one model to another. We have considered the case of a class of models referred as G2-MSSM \cite{Acharya:2008zi,Acharya:2008bk}. Here DM is represented by a system of two coannihilating species: the lightest neutralino and chargino (both wino-like) which are quasi degenerate in mass. The remaining spectum is at a very high mass scale, $O(10-100) \mbox{TeV}$ with the exception of the other gauginos, below the TeV scale, but non relevant for coannihilations. 
The issue of relaxing the condition of kinetic equilibrium have been approached in two steps. First of all we have verified the thermalization of the two species soon after production. This  depends on their energy loss rates due interactions with the SM particles of the thermal bath. Thermalization occurs essentially instantaneously due to the following facts: at temperatures higher than order of 1 GeV charginos are promptly converted into neutralinos with the latters losing energy very efficiently. At lower temperatures charginos are able to lose energy before being converted and guarantee the thermalization of the two species system as long as inelastic neutralino scatterings are efficient. This picture is summarized in fig.~(\ref{fig:energyloss}).
Once thermalization is accomplished, the evolution of the system until kinetic decoupling have been followed by writing two Boltzmann equations for the distribution functions of the two species. These equation can be integrated by expanding the collisonal operator in an analogous manner to \cite{Bringmann:2006mu,Bringmann:2009vf}:
\begin{eqnarray}
   \label{eq:nds}
   \frac{dn_{\chi^0}}{dt} + 3\,H\,n_{\chi^0} &=&  
    \left(  \widetilde{\Gamma}_{\chi^0 \leftrightarrow \chi^\pm} +\Gamma_{\chi^\pm}\right) \left[g_{\chi^0} n_{\chi^\pm} -{g}_{\chi^\pm} n_{\chi^0} \exp{\left(-\frac{\Delta m_\chi}{T}\right)}\right] \\ \nonumber
    &&  - \langle\sigma v\rangle_{\chi^0 \chi^0} \left[n_{\chi^0}^2 - (n_{\chi^0}^{eq})^2 \right] 
   -  \langle\sigma v\rangle_{\chi^0 \chi^\pm} \left[n_{\chi^0} n_{\chi^\pm} - n_{\chi^0}^{eq} n_{\chi^\pm}^{eq} \right]
\\ \nonumber\
   \frac{dn_{\chi^\pm}}{dt} + 3\,H\,n_{\chi^\pm} &=&
    \left( \widetilde{\Gamma}_{\chi^0 \leftrightarrow \chi^\pm} + \Gamma_{\chi^\pm} \right) \left[{g}_{\chi^\pm} n_{\chi^0} \exp{\left(-\frac{\Delta m_\chi}{T}\right)} - g_{\chi^0} n_{\chi^\pm}\right]
     - \langle\sigma v\rangle_{\chi^\pm \chi^\pm} \left[n_{\chi^\pm}^2 - (n_{\chi^\pm}^{eq})^2 \right] \\ \nonumber
   &&-  \langle\sigma v\rangle_{\chi^\pm \chi^0} \left[n_{\chi^\pm} n_{\chi^0} - n_{\chi^\pm}^{eq} n_{\chi^0}^{eq} \right]
     + \sum_i \frac{B_{X_i}}{m_{X_i}} \Gamma_{X_i} \rho_{X_i} \,. 
\label{eq:singlen}
\end{eqnarray}
with $\widetilde{\Gamma}_{\chi^0 \leftrightarrow \chi^\pm}$ and $\Gamma_{\chi^\pm}$, being, respectively the scattering and the decay rate. The last term in the second equation is the non-thermal production term and is defined in \cite{Arcadi:2011ev}.  The same formalism describes the dark matter distribution close to kinetic decoupling. Indeed, integrating the second moment of the distribution function it is possible to define the effective temperature $T_{\chi_0}$ (we define only one temperature being charginos always kept in equilibrium by electromagnetic interactions) which coincides with the thermal bath temperature until there is kinetic equilibrium while  scales as $T_{\chi_0} \propto T^2$ after kinetic decoupling. In the second panel of fig~(\ref{fig:Tsever}) we show the behaviour of  $T_{\chi_0}$ for different values of the reheating temperature. The impact of the decay of the cosmological moduli is twofold (fig.~(\ref{fig:Tsever})): the chargino decays tend to populate the system with neutralinos that are on average more energetic than for a thermal distribution, delaying the onset of the
regime $T_{{\chi}^{0}} \propto T^2$ and making the transition into this regime to be less sharp; at the same time, if $T_{\rm RH}$ is so low that reheating increases
significantly the expansion rate of the Universe $H$ at the time of kinetic decoupling  the departure from
$T_{{\chi}_{0}}=T$ tends to be anticipated.\\
In conclusion we point out that the developed formalism for the study of dark matter close to kinetic decoupling can be applied to other particle physics frameworks, also within standard cosmology.

\begin{thebibliography}{99}

\bibitem{Gondolo:2004sc}
  P.~Gondolo, J.~Edsjo, P.~Ullio, L.~Bergstrom, M.~Schelke and E.~A.~Baltz,
  JCAP {\bf 0407} (2004) 008
  [arXiv:astro-ph/0406204].
  
\bibitem{ArkaniHamed:2004yi}
  N.~Arkani-Hamed, S.~Dimopoulos, G.~F.~Giudice and A.~Romanino,
  Nucl.\ Phys.\  B {\bf 709} (2005) 3
  [arXiv:hep-ph/0409232].

\bibitem{Giudice:2004tc}
G.~F.~Giudice and A.~Romanino,
Nucl.\ Phys.\ B {\bf 699} (2004) 65
[arXiv:hep-ph/0406088].

\bibitem{Acharya:2008zi}
  B.~S.~Acharya, K.~Bobkov, G.~L.~Kane, J.~Shao and P.~Kumar,
  Phys.\ Rev.\  D {\bf 78} (2008) 065038
  [arXiv:0801.0478 [hep-ph]].
  
\bibitem{Acharya:2008bk}
  B.~S.~Acharya, P.~Kumar, K.~Bobkov, G.~Kane, J.~Shao and S.~Watson,
  JHEP {\bf 0806} (2008) 064
  [arXiv:0804.0863 [hep-ph]].
  
\bibitem{Bringmann:2006mu}
  T.~Bringmann and S.~Hofmann,
  JCAP {\bf 0407} (2007) 016
  [arXiv:hep-ph/0612238].
  
\bibitem{Bringmann:2009vf}
  T.~Bringmann,
  New J.\ Phys.\  {\bf 11} (2009) 105027
  [arXiv:0903.0189 [astro-ph.CO]].
  
\bibitem{Arcadi:2011ev}
  G.~Arcadi and P.~Ullio,
  Phys.\ Rev.\  D {\bf 84} (2011) 043520
  [arXiv:1104.3591 [hep-ph]].



\end{thebibliography}
\end{document}